\documentclass[runningheads]{svmult}
\usepackage{makeidx}   
\usepackage{subeqnar}  
\usepackage{physprbb}  
\makeindex             
\begin{document}

\title*{Cosmological Studies with Gamma-Ray Bursts}

\toctitle{Cosmological Studies with Gamma-Ray Bursts}
\titlerunning{Cosmological Studies with Gamma-Ray Bursts}

\author{Abraham Loeb}
\authorrunning{Avi Loeb}
\institute{Astronomy Department, Harvard University\\
Cambridge, MA 02138, USA}

\maketitle              

\begin{abstract}

Gamma-Ray Burst (GRB) explosions from the first generation of stars offer
an exciting opportunity to probe the epoch of reionization.  Clues about
how and when the intergalactic medium was ionized can be read off their UV
emission spectrum. Roughly a percent of all GRBs should be strongly
gravitationally lensed by intervening stars. A microlensed lightcurve can
be inverted to reconstruct the surface brightness profile of the GRB image
on the sky, with micro--arcsecond resolution.

\end{abstract}

\section{Introduction}

Since their discovery four decades ago, quasars have been used as powerful
lighthouses which probe the intervening universe out to high redshifts,
$z\sim 6$ (\cite{Hartwick,Fan}).  The spectra of almost all quasars show
strong emission lines of metals, indicating super-solar enrichment of the
emitting gas \cite{Hamann}. This implies that at least in the cores of
galaxies, formation of massive stars and their evolution to supernovae
preceded the observed quasar activity.  If Gamma-Ray Bursts (GRBs)
originate from the remnants of massive stars (such as neutron stars or
black holes), as seems likely based on recent estimates of their energy
output \cite{WKF,Freedman,Frail}, then they should exist at least out to
the same redshift as quasars.  Although GRBs are transient events, their
peak optical-UV flux can be as bright as that of quasars. Hence, GRBs
promise to be as useful as quasars in probing the high--redshift universe.

\begin{figure}[htbp]
\includegraphics{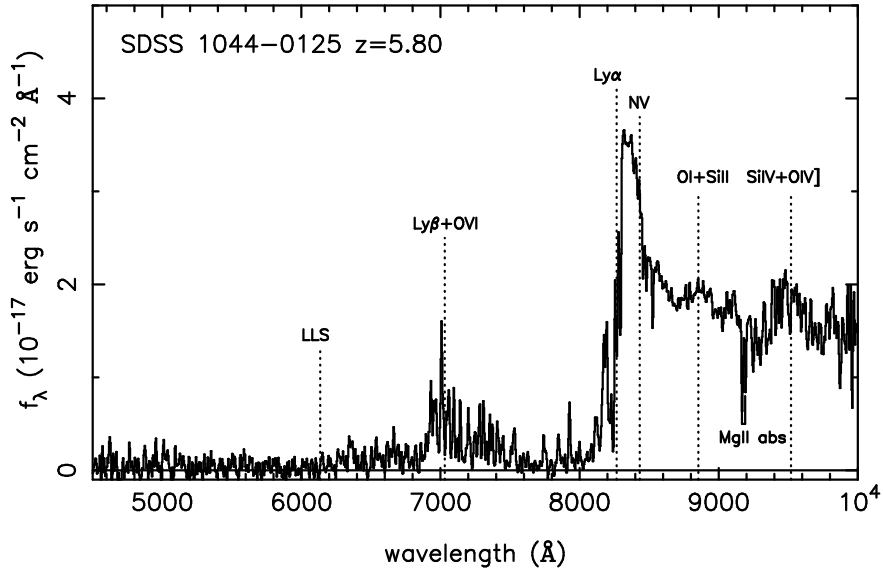}
\vspace{3.1in}
\caption{Optical spectrum of a $z=5.8$ quasar, discovered by the Sloan
Digital Sky Survey (Fan et al. 2000).  }
\label{fig1}
\end{figure}

Not much is known observationally about the universe in the redshift
interval $z=6$--$30$, when the first generation of galaxies condensed out
of the primordial gas left over from the Big Bang (see reviews
\cite{Barkana} and \cite{Loeb_Barkana}). Observations of the microwave
background anisotropies indicate that the cosmic gas became neutral at
$z\sim 1000$ and remained so at least down to $z\sim 30$ (see,
e.g. \cite{Wang}).  On the other hand, the existence of transmitted flux
shortward of the Ly$\alpha$ resonance in the spectrum of the
highest-redshift quasars and galaxies (see, for example,
Figure~\ref{fig1}), indicates that the intergalactic medium was reionized
to a level better than 99.9999\% by a redshift $z\sim 6$.  This follows
from the fact that the Ly$\alpha$ optical depth of the intergalactic medium
at high-redshifts ($z\gg1$) is \cite{Gunn},
\begin{equation}
\tau_{\alpha}={\pi e^2 f_\alpha \lambda_\alpha n_{HI}(z) \over m_e
cH(z)} \approx 6.45\times 10^5 x_{HI} \left({\Omega_bh\over
0.03}\right)\left({\Omega_m\over 0.3}\right)^{-1/2} \left({1+z\over
10}\right)^{3/2}, \label{eq:G-P}
\end{equation}
where $H\approx 100h~{\rm km~s^{-1}~Mpc^{-1}}\Omega_m^{1/2}(1+z)^{3/2}$ is
the Hubble parameter at the source redshift $z$, $f_\alpha=0.4162$ and
$\lambda_\alpha=1216$\AA\, are the oscillator strength and the wavelength
of the Ly$\alpha$ transition; $n_{HI}(z)$ is the average intergalactic
density of neutral hydrogen at the source redshift (assuming primordial
abundances); $\Omega_m$ and $\Omega_b$ are the present-day density
parameters of all matter and of baryons, respectively; and $x_{HI}$ is the
average fraction of neutral hydrogen. Modeling \cite{Fan} of the
transmitted flux in Figure~\ref{fig1} implies $\tau_{\alpha}<0.5$ or
$x_{HI}< 10^{-6}$, i.e., the low-density gas throughout the universe is
fully ionized at $z\sim 6$!

The question: {\it how and when was the universe reionized?}  defines a new
frontier in observational cosmology \cite{Loeb_Barkana}. The UV spectrum of
GRB afterglows can be used to probe the ionization and thermal state of the
intergalactic gas during the epoch of reionization, at redshifts $z\sim
7$--10 \cite{Miralda_Escude}.  The stretching of the temporal evolution of
GRB lightcurves by the cosmological redshift factor $(1+z)$, makes it
easier for an observer to react in time and take a spectrum of their
optical-UV emission at its peak.

Energy arguments suggest that reionization resulted from photoionization
rather than from collisional ionization \cite{Tegmark,Furlanetto}.  The
corresponding sources of UV photons were either stars or quasars.  Recent
simulations of the first generation of stars that formed out of the
primordial metal--free gas indicate that these stars were likely massive
\cite{Bromm,Abel}. If GRBs result from compact stellar remnants, such as
black holes or neutron stars, then the fraction of all stars that lead to
GRBs may have been higher at early cosmic times. This, however, is true
only if the GRB phenomena is triggered on a time scale much shorter than
the age of the universe at the corresponding redshift, which for $z\gg 1$
is $\sim 5.4\times 10^8~{\rm yr}~(h/0.7)^{-1}(\Omega_m/0.3)^{-1/2}
[(1+z)/10]^{-3/2}$. This condition may not hold, for example, for neutron
star binaries with an excessively long coalescence time.

In \S 2 we briefly discuss the expected properties of GRB afterglows at
high redshift. We then describe the use of distant GRBs for two studies:
(i) probing the intergalactic medium (IGM) during the epoch of reionization
(\S 3); and (ii) finding intervening stars at cosmological distances
through their gravitational lensing effect (\S 4).

\section{Properties of High-Redshift GRB Afterglows}

\begin{figure}[htbp]
\includegraphics{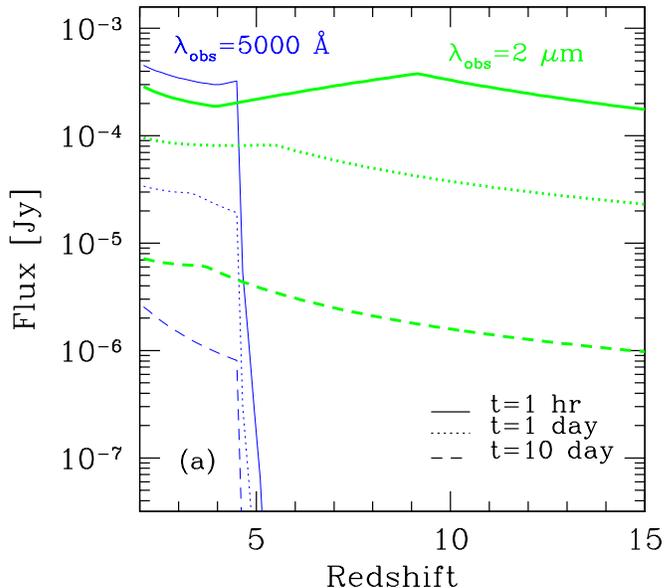}
\vspace{3.25in}
\caption{Theoretical expectation for the observed afterglow flux of a GRB
as a function of its redshift (from Ciardi \& Loeb 2000).  The curves refer
to an observed wavelength of $5000$ \AA (thin lines) and $2 \mu$m (thick
lines). Different line types refer to different observed times after the
GRB trigger, namely 1 hour (solid line), 1 day (dotted) and 10 days
(dashed). The $5000$\AA~ flux is strongly absorbed at $z>4.5$ by
intergalactic hydrogen. However, at infrared and radio wavelengths the
observed afterglow flux shows only a mild dependence on the source
redshift.}
\label{fig2}
\end{figure}

Young (days to weeks old) GRBs outshine their host galaxies in the optical
regime. In the standard hierarchical model of galaxy formation, the
characteristic mass and hence optical luminosity of galaxies and quasars
declines with increasing redshift \cite{Haiman,Stern,Barkana}. Hence, GRBs
should become easier to observe than galaxies or quasars at increasing
redshift. Similarly to quasars, GRB afterglows possess broad-band spectra
which extend into the rest-frame UV and can probe the ionization state and
metallicity of the IGM out to the epoch when it was reionized at a redshift
$z\sim 7$--$10$ \cite{Loeb_Barkana}.  Lamb \& Reichart \cite{Lamb} have
extrapolated the observed $\gamma$-ray and afterglow spectra of known GRBs
to high redshifts and emphasized the important role that their detection
might play in probing the IGM. Simple scaling of the long-wavelength
spectra and temporal evolution of afterglows with redshift implies that at
a fixed time lag after the GRB trigger in the observer's frame, there is
only a mild change in the {\it observed} flux at infrared or radio
wavelengths as the GRB redshift increases. Ciardi \& Loeb \cite{Ciardi}
demonstrated this behaviour using a detailed extrapolation of the GRB
fireball solution into the non-relativistic regime (see the 2$\mu$m curves
in Figure~\ref{fig2}). Despite the strong increase of the luminosity
distance with redshift, the observed flux for a given observed age is
almost independent of redshift in part because of the special spectrum of
GRB afterglows (see Figure~\ref{fig4}), but mainly because afterglows are
brighter at earlier times and a given observed time refers to an earlier
intrinsic time in the source frame as the source redshift increases. The
mild dependence of the long-wavelength ($\lambda_{\rm obs}>1\mu$m) flux on
redshift differes from other high-redshift sources such as galaxies or
quasars, which fade rapidly with increasing redshift
\cite{Haiman,Stern,Barkana}. Hence, GRBs provide exceptional lighthouses
for probing the universe at $z=6$--30, at the epoch when the first stars
had formed.

Assuming that the GRB rate is proportional to the star formation rate and
that the characteristic energy output of GRBs is $\sim 10^{52}~{\rm ergs}$,
Ciardi \& Loeb \cite{Ciardi} predicted that there are always $\sim 15$ GRBs
from redshifts $z> 5$ across the sky which are brighter than $\sim 100$ nJy
at an observed wavelength of $\sim 2\mu$m.  The infrared spectrum of these
sources could be taken with future telescopes such as the {\it Next
Generation Space Telescope} (planned for launch in 2009; see
http://ngst.gsfc.nasa.gov/), as a follow-up on their early X-ray
localization with the {\it Swift} satellite (planned for launch in 2003;
see http://swift.sonoma.edu/).

The redshifts of GRB afterglows can be estimated photometrically from
either the Lyman limit or Ly$\alpha$ troughs in their spectra. At low
redshifts, the question of whether the Lyman limit or Ly$\alpha$ trough
interpretation applies depends on the absorption properties of the host
galaxy. If the GRB originates from within the disk of a star--forming
galaxy, then the afterglow spectrum will likely show a damped Ly$\alpha$
trough. At $z> 6$ the Ly$\alpha$ trough would inevitably exist since the
intergalactic Ly$\alpha$ opacity is $> 90\%$ (see Figure 13 in
\cite{Stern}). Interestingly, an absorption feature in the afterglow
spectrum which is due to the neutral hydrogen within a molecular cloud or
the disk of the host galaxy, is likely to be time-dependent due to the
ionization caused by the UV illumination of the afterglow itself along the
line-of-sight \cite{Perna}.

So far, there have been two claims for high-redshift GRBs. Fruchter
\cite{Fruchter} argued that the photometry of GRB 980329 is consistent with
a Ly$\alpha$ trough due to IGM absorption at $z\sim 5$.  Anderson et
al. \cite{Anderson} inferred a redshift of $z=4.5$ for GRB 000131 based on
a crude optical spectrum that was taken by the VLT a few days after the GRB
trigger. {\it These cases emphasize the need for a coordinated observing
program that will alert 10-meter class telescopes to take a spectrum of an
afterglow about a day after the GRB trigger, based on a photometric
assessment (obtained with a smaller telecope using the Lyman limit or
Ly$\alpha$ troughs) that the GRB may originate at a high redshift.}

In the following two sections, we use two examples to illustrate the
usefulness of GRB afterglows for cosmological studies.

\begin{figure}[htbp]
\includegraphics{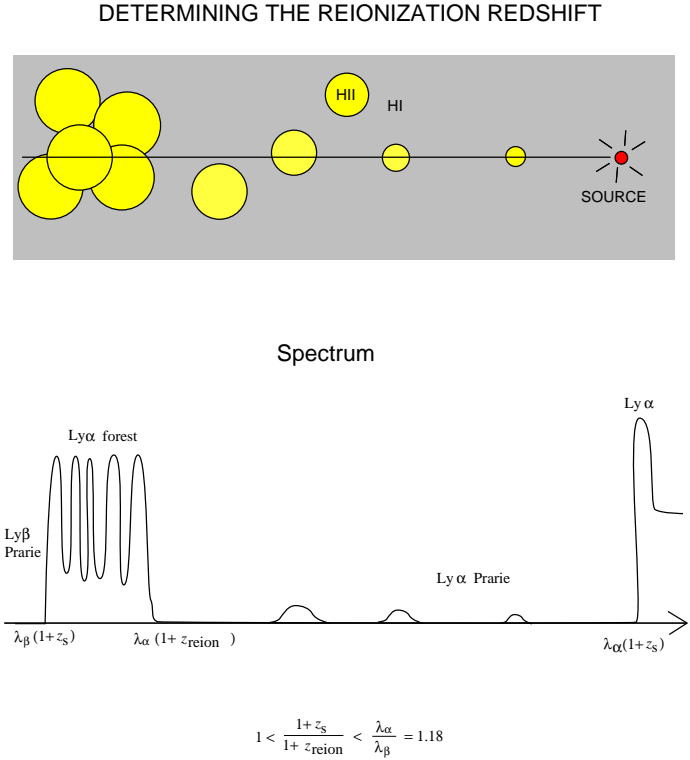}
\vspace{4.2in}
\caption{ Sketch of the expected spectrum of a source at a redshift $z_{s}$
slightly above the reionization redshift $z_{\rm reion}$. The transmitted
flux due to HII bubbles in the pre-reionization era, and the Ly$\alpha$
forest in the post-reionization era, are exaggerated for illustration.  }
\label{fig3}
\end{figure}

\section{Probing the Intergalactic Medium}

The UV spectra of GRB afterglows can be used to measure the evolution of
the neutral intergalactic hydrogen with redshift. Figure~\ref{fig3}
illustrates schematically the expected absorption just beyond the
reionization redshift. Resonant scattering suppresses the spectrum at all
wavelengths corresponding to the Ly$\alpha$ resonance prior to
reionization.  Since the Ly$\alpha$ cross-section is very large, any
transmitted flux prior to reionization reflects a large volume of ionized
hydrogen along the line-of-sight. If the GRB is located at a redshift
larger by $>18\%$ than the reionization redshift, then the Ly$\alpha$ and
the Ly$\beta$ troughs will overlap. Unlike quasars, GRBs do not ionize the
IGM around them; their limited energy supply $\sim 10^{52}~{\rm ergs}$
\cite{WKF,Freedman,Frail} can ionize only $\sim 4\times 10^5M_\odot$ of
neutral hydrogen within their host galaxy.

Quasar spectra indicate the existence of an IGM metallicity which is a
fraction of a percent of the solar value \cite{Ellison}.  The metals were
likely dispersed into the IGM through outflows from galaxies, driven by
either supernova or quasar winds \cite{Barkana,Furlanetto}. Detection of
metal absorption lines in the spectrum of GRB afterglows, produced either
in the IGM or the host galaxy of the GRB, can help unravel the evolution of
the IGM metallicity with redshift and its link to the evolution of
galaxies.

\section{Cosmological Microlensing of Gamma-Ray Bursts}

Loeb \& Perna \cite{Loeb_Perna} noted the coincidence between the angular
size of a solar-mass lens at a cosmological distance and the
micro-arcsecond size of the image of a GRB afterglow. They therefore
suggested that microlensing by stars can be used to resolve the
photospheres of GRB fireballs at cosmological distances.  (Alternative
methods, such as radio scintillations, only provide a constraint on the
radio afterglow image size \cite{Goodman,WKF} but do not reveal its
detailed surface brightness distribution, because of uncertainties in the
scattering properties of the Galactic interstellar medium.)

\begin{figure}[htbp]
\includegraphics{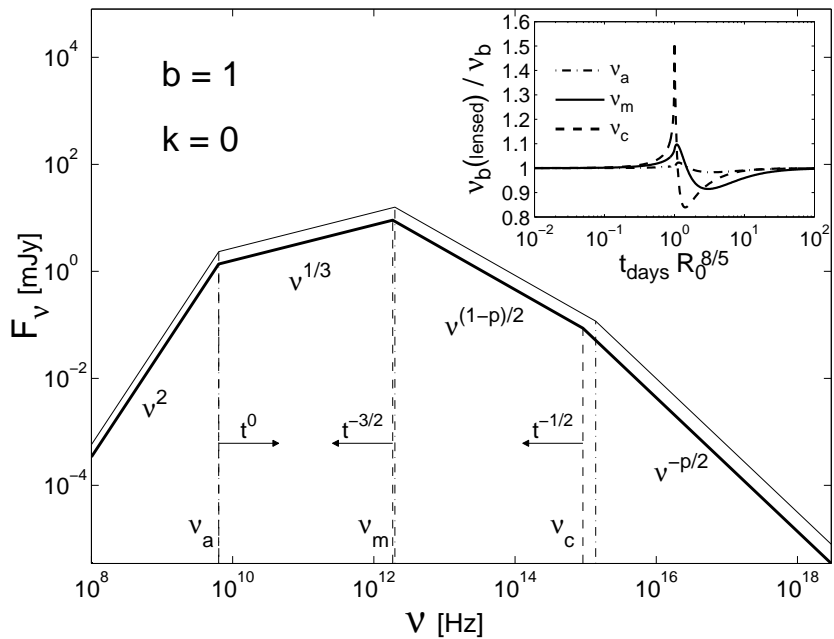}
\vspace{3.6in}
\caption{A typical broken power-law spectrum of a GRB afterglow at a
redshift $z=1$ (from Granot \& Loeb 2001). The observed flux density,
$F_\nu$, as a function of frequency, $\nu$, is shown by the boldface solid
line at an observed time $t_{\rm days}=1$ for an explosion with a total
energy output of $10^{52}~{\rm ergs}$ in a uniform interstellar medium
($k=0$) with a hydrogen density of $1~{\rm cm^{-3}}$, and post-shock energy
fractions in accelerated electrons and magnetic field of $\epsilon_e=0.1$
and $\epsilon_B=0.03$, respectively.  The thin solid line shows the same
spectrum when it is microlensed by an intervening star with an impact
parameter equal to the Einstein angle and $R_0\equiv [\theta_s(1~{\rm
day})/\theta_{\rm E}]=1$. The insert shows the excess evolution of the
break frequencies $\nu_{\rm b}=\nu_a,~\nu_m$ and $\nu_c$ (normalized by
their unlensed values) due to microlensing. }
\label{fig4}
\end{figure}

The fireball of a GRB afterglow is predicted to appear on the sky as a ring
(in the optical band) or a disk (at low radio frequencies) which expands
laterally at a superluminal speed, $\sim \Gamma c$, where $\Gamma\gg1$ is
the Lorentz factor of the relativistic blast wave which emits the afterglow
radiation \cite{Waxman,Sari,Panaitescu,Granot}.  For a spherical explosion
into a constant density medium (such as the interstellar medium), the
physical radius of the afterglow image is of order the fireball radius over
$\Gamma$, or more precisely \cite{Granot}
\begin{equation}
R_{\rm s}= 3.9\times 10^{16}\left({E_{52}\over n_1}\right)^{1/8}
\left({t_{\rm days}\over 1+z}\right)^{5/8}~{\rm cm},
\label{eq:r_s}
\end{equation}
where $E_{52}$ is the hydrodynamic energy output of the GRB explosion in
units of $10^{52}~{\rm ergs}$, $n_1$ is the ambient medium density in units
of $1~{\rm cm^{-3}}$, and $t_{\rm days}$ is the observed time in days.  At
a cosmological redshift $z$, this radius of the GRB image occupies an angle
$\theta_{\rm s}=R_{\rm s}/D_A$, where $D_A(z)$ is the angular diameter
distance at the GRB redshift, $z$. For the typical cosmological distance,
$D_A\sim 10^{28}~{\rm cm}$, the angular size is of order a micro-arcsecond
($\mu$as).  Coincidentally, this image size is comparable to the Einstein
angle of a solar mass lens at a cosmological distance,
\begin{equation}
\theta_{\rm E}=\left({4GM_{\rm lens}\over c^2 D}\right)^{1/2}= 1.6
\left({M_{\rm lens}\over 1 M_\odot}\right)^{1/2}\left({D\over 10^{28}~{\rm
cm}}\right)^{-1/2}~\mu{\rm as},
\label{eq:1}
\end{equation}
where $M_{\rm lens}$ is the lens mass, and $D\equiv {D_{\rm os}D_{\rm ol}/
D_{\rm ls}}$ is the ratio of the angular-diameter distances between the
observer and the source, the observer and the lens, and the lens and the
source \cite{Schneider}.  Loeb \& Perna (1998) showed that because the ring
expands laterally faster than the speed of light, the duration of the
microlensing event is only a few days rather than tens of years, as is the
case for more typical astrophysical sources which move at a few hundred
${\rm km~s^{-1}}$ or $\sim 10^{-3}c$.

The microlensing lightcurve goes through three phases: (i) constant
magnification at early times, when the source is much smaller than the
source-lens angular separation; (ii) peak magnification when the ring-like
image of the GRB first intersects the lens center on the sky; and (iii)
fading magnification as the source expands to larger radii.

\begin{figure}[htbp]
\includegraphics{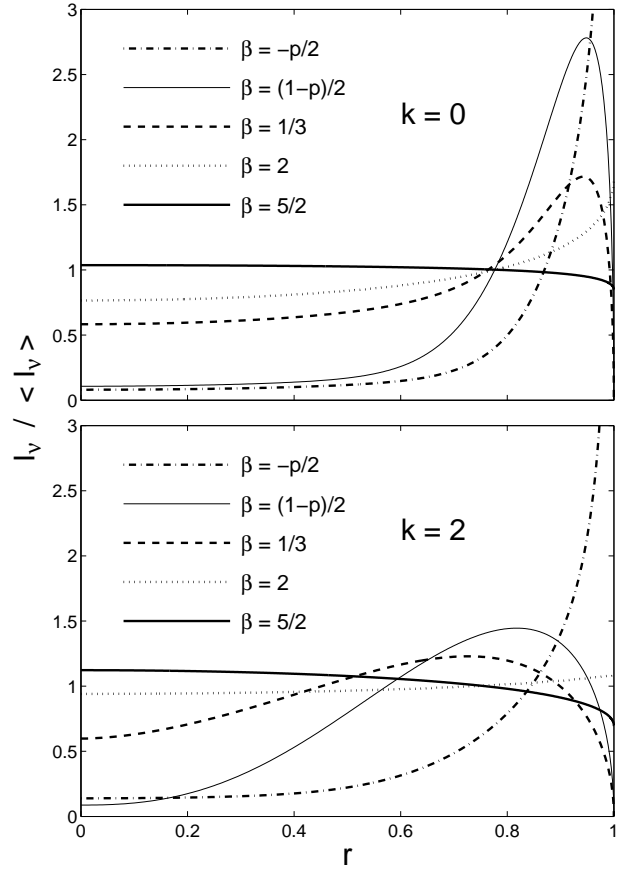}
\vspace{4.8in}
\caption{The surface brightness, normalized by its average value, as a
function of the normalized radius, $r$, from the center of a GRB afterglow
image (where $r=0$ at the center and $r=1$ at the outer edge). The image
profile changes considerably between different power-law segments of the
afterglow spectrum, $F_{\nu}\propto\nu^{\beta}$ (see
Figure~\ref{fig4}). There is also a strong dependence on the power--law
index of the radial density profile of the external medium around the
source, $\rho\propto R^{-k}$ (taken from Granot \& Loeb 2001).  }
\label{fig5}
\end{figure}

Granot \& Loeb \cite{Granot_Loeb} calculated the radial surface brightness
profile (SBP) of the image of a Gamma-Ray-Burst (GRB) afterglow as a
function of frequency and ambient medium properties, and inferred the
corresponding magnification lightcurves due to microlensing by an
intervening star. The afterglow spectrum consists of several power-law
segments separated by breaks, as illustrated by Figure~\ref{fig4}. The
image profile changes considerably across each of the spectral breaks, as
shown in Figure~\ref{fig5}. It also depends on the power--law index, $k$,
of the radial density profile of the ambient medium into which the GRB
fireball propagates.  Gaudi \& Loeb \cite{Gaudi_Loeb} have shown that
intensive monitoring of a microlensed afterglow lightcurve can be used to
reconstruct the parameters of the fireball and its environment.  The
dependence of the afterglow image on frequency offers a fingerprint that
can be used to identify a microlensing event and distinguish it from
alternative interpretations.  It can also be used to constrain the
relativistic dynamics of the fireball and the properties of its gaseous
environment. At the highest frequencies, the divergence of the surface
brightness near the edge of the afterglow image ($r=1$ in
Figure~\ref{fig5}) depends on the thickness of the emitting layer behind
the relativistic shock front, which is affected by the length scale
required for particle acceleration and magnetic field amplification behind
the shock\cite{Medvedev,Gruzinov}.

Ioka \& Nakamura \cite{Ioka} considered the more complicated case where the
explosion is collimated and centered around the viewing axis. More general
orientations that violate circular symmetry need to be considered in the
future.

\begin{figure}[htbp]
\includegraphics{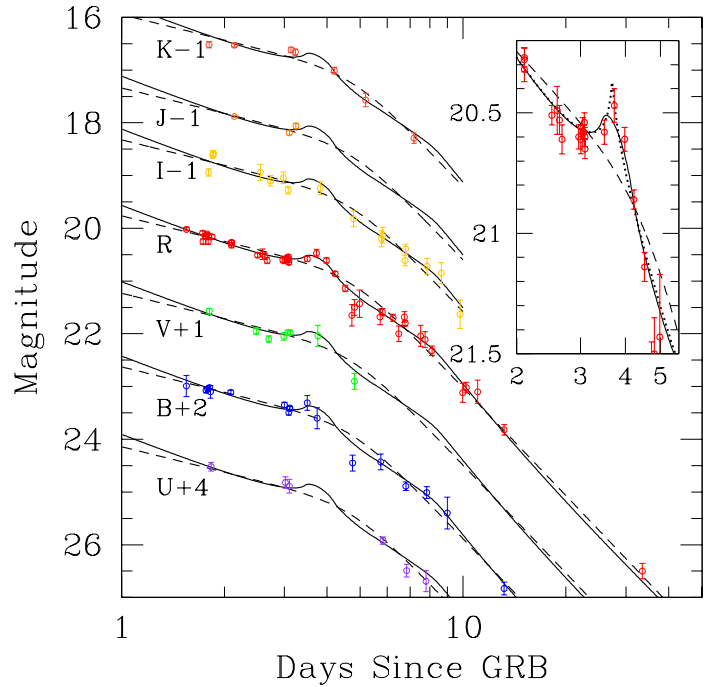}
\vspace{3.9in}
\caption{ $UBVRIJK$ photometry of GRB 000301C as a function of time in days
from the GRB trigger (from Garnavich et al. 2000; Gaudi et al.  2001).
Data points have been offset by the indicated amount for clarity.  The
dashed line is the best-fit smooth, double power-law lightcurve (with no
lensing), while the solid line is the overall best-fit microlensing model,
where the SBP has been determined from direct inversion.  The inset shows
the $R$-band data only.  The dotted line is the best-fit microlensing model
with theoretically calculated SBP, for $k=0$ and $\nu>\nu_c$.  }
\label{fig6}
\end{figure}

\subsection{GRB 000301C}

Garnavich, Loeb, \& Stanek \cite{Garnavich} have reported the possible
detection of a microlensing magnification feature in the optical-infrared
light curve of GRB 000301C (see Figure~\ref{fig6}). The achromatic
transient feature is well fitted by a microlensing event of a $0.5 M_\odot$
lens separated by an Einstein angle from the source center, and resembles
the prediction of Loeb \& Perna \cite{Loeb_Perna} for a ring-like source
image with a narrow fractional width ($\sim 10\%$). Alternative
interpretations relate the transient achromatic brightening to a higher
density clump into which the fireball propagates \cite{Berger}, or to a
refreshment of the decelerating shock either by a shell which catches up
with it from behind or by continuous energy injection from the source
\cite{Zhang}. However, the microlensing model has a smaller number of free
parameters. If with better data, a future event will show the generic
temporal and spectral characteristics of a microlensing event, then these
alternative interpretations will be much less viable.  A galaxy
2$^{\prime\prime}$ from GRB~000301C might be the host of the stellar lens,
but current data provides only an upper-limit on its surface brightness at
the GRB position. The existence of an intervening galaxy increases the
probability for microlensing over that of a random line-of-sight.

\begin{figure}[htbp]
\includegraphics{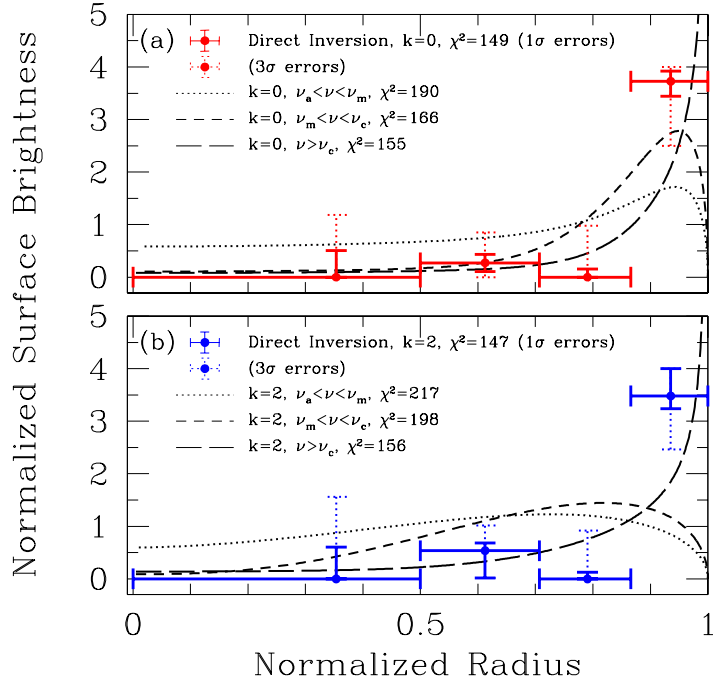}
\vspace{3.9in}
\caption{Fitting GRB 000301C with different SBPs as a function of
normalized radius (taken from Gaudi et al. 2001).  The points are the SBPs
determined from direct inversion, with $1\sigma$ errors (solid) and
$3\sigma$ errors (dotted).  The curves are theoretically calculated SBPs
for various frequency regimes (see Figure~\ref{fig5}).  (a) Uniform
external medium, $k=0$.  (b) Stellar wind environment, $k=2$. The number of
degrees of freedom is 92 for the direct inversion points and 89 for the
curves.}
\label{fig7}
\end{figure}

Gaudi, Granot, \& Loeb \cite{GGL} have shown that direct inversion of the
observed light curve for GRB 000301C yields a surface brightness profile
(SBP) of the afterglow image which is strongly limb-brightened, as expected
theoretically (see Figure~\ref{fig7}).

Obviously, realistic lens systems could be more complicated due to the
external shear of the host galaxy or a binary companion.  Mao \& Loeb
\cite{Mao} calculated the magnification light curves in these cases, and
found that binary lenses may produce multiple peaks of magnification.  They
also demonstrated that {\it all} afterglows are likely to show variability
at the level of a few percent about a year following the explosion, due to
stars which are separated by tens of Einstein angles from the
line-of-sight.

{\it What is the probability for microlensing?} If the lenses are not
strongly clustered so that their cross-sections overlap on the sky, then
the probability for having an intervening lens star at a projected angular
separation $\theta$ from a source at a redshift $z\sim 2$ is~~$\sim 0.3
\Omega_\star (\theta/\theta_E)^2$ \cite{Press,Blaes,Nemiroff,Nemiroff2},
where $\Omega_\star$ is the cosmological density parameter of stars.  The
value of $\Omega_\star$ is bounded between the density of the luminous
stars in galaxies and the total baryonic density as inferred from Big Bang
nucleosynthesis, $7\times 10^{-3} < \Omega_\star< 5\times 10^{-2}$
\cite{Fukugita}. Hence, {\it all} GRB afterglows should show evidence for
events with $\theta < 30\theta_E$, for which microlensing provides a small
perturbation to the light curve\cite{Mao}. (This crude estimate ignores the
need to subtract those stars which are embedded in the dense central
regions of galaxies, where macrolensing dominates and the microlensing
optical depth is of order unity.) However, only one out of roughly a
hundred afterglows is expected to be strongly microlensed with an impact
parameter smaller than the Einstein angle. Indeed, Koopmans \& Wambsganss
\cite{Koopmans} have found that the `a posteriori' probability for the
observed microlensing event of GRB 000301C along a random line-of-sight is
between 0.7--2.7\% if 20-100\% of the dark matter is in compact objects.

Microlensing events are rare but precious.  Detailed monitoring of a few
strong microlensing events among the hundreds of afterglows detected per
year by the forthcoming Swift satellite, could be used to constrain the
environment and the dynamics of relativistic GRB fireballs, as well as
their magnetic structure and particle acceleration process.

\bigskip
\noindent{\bf Acknowledgements}

\noindent
I thank all my collaborators on this subject: the young theorists with whom
I studied the above topics -- Rennan Barkana, Benedetta Ciardi, Scott
Gaudi, Jonathan Granot, Zoltan Haiman, Shude Mao, Misha Medvedev, and
Rosalba Perna; and the observers -- Peter Garnavitch and Kris Stanek, who
introduced me to the data on GRB 000301C.

\vfil\eject

\end{document}